\documentclass[]{emulateapj}
\usepackage{color}
\usepackage{amssymb}
\usepackage{amsmath}
\begin{document}

\title{ An Extreme Starburst    in Close Proximity to the Central Galaxy  of a  Rich Galaxy Cluster at $\lowercase{z}= 1.7$}

\author{Tracy Webb\altaffilmark{1}}

\author{Allison Noble\altaffilmark{2}}
\author{Andrew DeGroot\altaffilmark{3}}
\author{Gillian Wilson\altaffilmark{3}}
\author{Adam Muzzin\altaffilmark{4}}
\author{Nina Bonaventura\altaffilmark{1}}
\author{Mike Cooper\altaffilmark{5}}
\author{Anna Delahaye\altaffilmark{1}}
\author{Ryan Foltz\altaffilmark{3}}
\author{Chris Lidman\altaffilmark{6}}
\author{Jason Surace\altaffilmark{7}}
\author{H.K.C. Yee\altaffilmark{2}}
\author{Scott Chapman\altaffilmark{4,8}}
\author{Loretta Dunne\altaffilmark{9,10}}
\author{James Geach\altaffilmark{11}}
\author{Brian Hayden\altaffilmark{12}}
\author{Hendrik Hildebrandt\altaffilmark{13}}
\author{Jiasheng Huang\altaffilmark{14,15,16}}
\author{Alexandra Pope\altaffilmark{17}}
\author{Matthew W.L. Smith\altaffilmark{18}}
\author{Saul Perlmutter\altaffilmark{12,19}}
\author{Alex Tudorica\altaffilmark{13}}

\altaffiltext{1}{Department of Physics, McGill University, 3600 rue University, Montr\'eal, Qu\'ebec, H3P 1T3, Canada}
\altaffiltext{2}{Department of Astronomy \& Astrophysics, University of Toronto, 50 St. George St., Toronto, ON, M5S 3H4, Canada}
\altaffiltext{3}{Department of Physics \& Astronomy, University of California, Riverside, 900 University Avenue, Riverside, CA 92521, USA}
\altaffiltext{4}{Institute of Astronomy, University of Cambridge, Madingley Road, Cambridge, CB3 0HA, UK}
\altaffiltext{5}{center for Galaxy Evolution, Department of Physics and Astronomy, University of California, Irvine, 4129 Frederick Reines Hall, Irvine, CA 92697 USA}
\altaffiltext{6}{Australian Astronomical Observatory, P.O. Box 915, North Ryde, NSW 1670, Australia}
\altaffiltext{7}{Spitzer Space Science center, California Institute of Technology, Pasadena, CA  91125 USA}
\altaffiltext{8}{Department of Physics and Atmospheric Science, Dalhousie University, 6310 Coburg Road, Halifax, NS B3H 2R4, Canada}
\altaffiltext{9}{Department of Physics and Astronomy, University of Canterbury, Private Bag 4800, Christchurch 8140, New Zealand}
\altaffiltext{10}{Insitute for Astronomy, University of Edinburgh, Royal Observatory, Blackford Hill, Edinburgh, EH9 3HJ, UK}
\altaffiltext{11}{Center for Astrophysics Research, Science \& Technology Research Institute, University of Hertfordshire, Hatfield AL10 9AB, UK}
\altaffiltext{12}{Physics Division, Lawrence Berkeley National Laboratory, 1 Cyclotron Road, Berkeley, CA 94720, USA}
\altaffiltext{13}{Argelander-Institute fur Astronomie, Auf dem Hugel 71, D-53121 Bonn, Germany}
\altaffiltext{14}{National Astronomical Observatories of China, Chinese Academy of Sciences, Beijing 100012, China}
\altaffiltext{15}{China-Chile Joint Center for Astronomy, Chinese Academy of Sciences, Camino El Observatory, 1515, Las Condes, Santiago, Chile}
\altaffiltext{16}{Harvard-Smithsonian Center for Astrophysics, 60 Garden St., Cambridge, MA 02138, USA}
\altaffiltext{17}{Department of Astronomy, University of Massachusetts, 710 North Pleasant Street, Amherst, MA 011003, USA}
\altaffiltext{18}{School of Physics and Astronomy, Cardiff University, Queens Buildings, The Parade, Cardiff, Wales CF24 3AA, UK} 
\altaffiltext{19}{Department of Physics, University of California Berkeley, 366 LeConte Hall MC 7300, Berkeley, CA 94720-7300, USA} 

\begin{abstract}
We have discovered  an optically rich galaxy cluster at $z=1.7089$ with star formation  occurring in close proximity to the   central galaxy.  The system, SpARCS104922.6+564032.5, was detected within  the {\it Spitzer} Adaptation of the red-sequence Cluster Survey, (SpARCS),  and confirmed through Keck-MOSFIRE spectroscopy.  The rest-frame optical richness  of $N_\mathrm{gal}$(500kpc) = 30$\pm$8 implies a  total halo mass, within 500kpc,  of $\sim$ 3.8$\pm1.2 \times$10$^{14}$ $M_\odot$, comparable to  other clusters at or above this redshift.  There is a wealth of ancillary data available, including Canada-France-Hawaii Telescope optical, UKIRT-K, {\it Spitzer}-IRAC/MIPS, and {\it Herschel}-SPIRE.   This work adds   submillimeter imaging with the  SCUBA2 camera on the James Clerk Maxwell Telescope and near-infrared imaging with the Hubble Space Telescope (HST).  The mid/far-infrared  (M/FIR) data detect an Ultra-luminous Infrared Galaxy spatially coincident with the central galaxy, with $L_\mathrm{IR}$ = 6.2$\pm0.9 \times$10$^{12}$ $L_\odot$.  The detection of polycyclic aromatic hydrocarbons (PAHs) at $z=1.7$  in a {\it Spitzer}-IRS spectrum of the source implies the FIR luminosity is dominated by star formation  (an Active Galactic Nucleus contribution of 20\%) with a  rate of $\sim$860$\pm$130$M_\odot$ yr$^{-1}$.  The optical  source corresponding to the IR emission is likely  a chain of   of $>$ 10 individual clumps  arranged as ``beads on a string'' over a linear scale of 66 kpc. Its morphology and proximity to the Brightest Cluster Galaxy imply a gas-rich interaction at the center of the cluster  triggered the star formation. This system indicates that wet  mergers  may be an important process in forming the stellar mass of BCGs at early times.

\end{abstract}

\keywords{galaxies: interactions, galaxies, starburst, galaxies: evolution, galaxies: clusters: general, infrared: galaxies }
\maketitle

\section{Introduction}

The most massive galaxies in the local universe reside at the centers of galaxy clusters.  Often called Brightest Cluster Galaxies  (BCGs), they exhibit, as a class, highly uniform properties such as luminosity, metallicity and surface brightness profiles, and are distinct from the general galaxy population \citep{oemler76,dressler78,schombert86}.      These facts suggest formation processes which are themselves distinct from those which dominate in massive galaxies   elsewhere in the cluster or in the field. The mass growth of BCGs  is likely linked to the overall physics of hierarchical structure formation on galaxy cluster scales, including the fundamental processes of gas cooling, star formation, energy feedback and galaxy mergers at the centers of giant dark matter halos.

Within the  current suite of Semi-Analytic Models (SAMs) BCG evolution follows a relatively simple prescription \citep{kauffmann00,springel01,springel05,delucia04,delucia06,croton06}.  At very early times ($z>5$) cooling flows dominate the stellar mass assembly at the centers of clusters, but these are rapidly suppressed  through active galactic nuclei (AGN) feedback \citep{bower06,delucia07,guo10}.   BCG growth then continues over time through the accretion of smaller systems, with little additional star formation. In this picture most of the stellar mass is formed at very high redshifts ($z\sim$ 5) within multiple galaxies through low levels of star formation, and the BCG slowly acquires its identity through the conglomeration of previously assembled pieces.  This is supported by direct measurements of the mass growth of BCGs below $z\sim$1 \citep{lidman12,lidman13}.   

Still, there are observational examples of moderate redshift BCGs ($z<0.6$) that present signs of  high star formation rates (SFRs) (10-1000$M_\odot$yr$^{-1}$) and/or contain  large amounts of molecular gas  \citep{allen92,mcnamara92,johnstone98,edge01,wilman06,edwards07,odea10,hicks10,donahue11,rawle12,mcdonald12}.  Such active BCGs are rare  and have primarily been seen in X-ray-selected cool-core clusters. In many cases the SFRs are correlated with the gas cooling time.  The paucity of these systems at low/moderate redshifts indicates that explosive star formation, fed by the  cooling of halo gas,  is not an important contributor to BCG mass growth, relative to the process of dry merging.


 
While it is established that a large fraction (50\%; Lidman et al. 2013) of the stellar mass of BCGs is already in place at $z\sim$ 1, it is not yet determined  if this mass was assembled through processes similar to those at lower redshift ({\it i.e.}, dry merging) or if  
a different and perhaps more extreme phase of growth,  such as that seen in some of the cool core clusters, becomes more common at earlier epochs.   There are now several  detections of clusters  and massive groups at redshifts beyond $z\sim$ 1  \citep{andreon09,muzzin09,wilson09,stanford12,zeimann12,gobat13,muzzin13,yuan14,bayliss14,tozzi15} and these systems offer the opportunity to study the build-up of the central stellar mass in detail. Indeed many of these systems exhibit an increase in the star formation activity within several hundred kpc of their centers, compared to lower redshifts   and, in some cases, the field \citep{tran10,brodwin13,alberts14,santos15}.  While studies of their BCGs, in particular, have not been undertaken, these  results indicate that the cores of galaxy clusters may be active places beyond $z\sim$ 1.5. 

Here we report the discovery of  an optically rich cluster at $z=$1.7089, that we label SpARCS104922.6+564032.5 (hererafter SpARCS1049+56). This system is distinguished by its extremely active BCG  which  is observed in an important phase of evolution. We appear to be  witnessing the stellar mass build-up of the central BCG through a wet galaxy merger, a process which has, heretofore, not been thought to  play an important role in BCG growth.    In this paper we outline the discovery and characterization of this interesting system. In Section 2 we describe the data and observations; in Section 3 we present the analysis of the cluster and BCG properties; and in Section 4 we discuss the implication of these first results.  We use  standard cosmology throughout ($\Omega_\mathrm{o}$ = 1, $\Omega_\Lambda$ = 0.7, $H_\mathrm{o}$ = 75 km s$^{-1}$Mpc$^{-1}$), and AB magnitudes.  






\section{Data and Observations}

The data used in this work were obtained with a broad range of facilities. Some were retrieved from  public archives while others were newly acquired by our team.  
Here we outline three categories of observations. In Section 2.1 we list the  wide-field imaging associated with large surveys of the {\it Spitzer}- Wide-area InfraRed Extragalactic (SWIRE) Lockman field, where the cluster is located.  In section 2.2  we describe  follow-up imaging with SCUBA2 on the James Clerk Maxwell Telescope (JCMT)  and WFC3 on the Hubble Space Telescope (HST) taken by our team subsequent to the cluster discovery (described in Section 3.1).  These data are focused on the location of the cluster and do not cover the entire Lockman field.  In Section 2.3 we present the  spectroscopic observations obtained by others before the discovery of the cluster and by ourselves following the discovery with Keck-MOSFIRE.  

\subsection{Existing Wide-field Imaging of the Lockman-SWIRE Field}

\subsubsection{The Spitzer Adaptation of the Red-sequence Cluster Survey:} The cluster was detected within the  {\it Spitzer} Adaptation of the red-sequence Cluster Survey (SpARCS) coverage of the Lockman Hole.   
SpARCS is a near-infrared survey for galaxy clusters beyond $z\sim$ 1 \citep{muzzin09,wilson09,demarco10}.  It combines deep $z'-$band imaging (5${\sigma}$ $ z' = $ 24.2 AB with the Canada-France-Hawaii Telescope; CFHT) with the SWIRE Survey \citep{lonsdale03} at 3.6$\mu$m,   to identify over-densities of red-sequence galaxies at high redshift.  A thorough discussion of the $z'$ data analysis,  $z'-3.6\mu$m cluster finding algorithm and the SpARCS cluster catalog will be presented in a future paper by A. Muzzin et al. (2015, in preparation), though we outline the method in Section 3.1.

\subsubsection{The Spitzer Space Telescope:} The public SWIRE Legacy Survey consists of imaging at 3.6, 4.5, 5.8 and 8.0$\mu$m with IRAC and 24, 70, 160$\mu$m with MIPS \citep{lonsdale03} \footnote{$http://$swire.ipac.caltech.edu$/$swire$/$astronomers$/$publications$/$ SWIRE2$\_$doc$\_$083105.pdf}.  SpARCS1049+56  lies near the edge of the survey and is not covered  by the SWIRE ancillary optical data for this field.   
Deeper ($\sim$ 2$\mu$Jy) 3.6/4.5$\mu$m imaging of this area has been obtained during the warm-{\it Spitzer} mission by the {\it Spitzer} Extragalactic Representative Volume Survey \citep{mauduit12}, but it is not used here.

\subsubsection{The Herschel Multi-tier Extragalactic Survey:} The Lockman-SWIRE field is also part of the {\it Herschel}\thanks{{\it Herschel is an ESA space observatory with science instruments provided by European-led Principal Investigator consortia and with important participation from NASA}} Mid-Infrared Extragalactic Survey (HerMES; Oliver et al. 2012).  The primary data for this survey are SPIRE 250/350/500$\mu$m imaging \citep{griffin10}, with additional PACS data taken in parallel mode.  PACS data products have not yet been released by the HerMES team and therefore we downloaded the original data from the {\it Herschel} Science Archive. We produced  maps at 100 and 160$\mu$m  following the technique developed in \citet{ibar10}.

\subsubsection{The UK Infrared Deep Sky Survey:} This field was imaged in the near-infrared as part of the UK Infrared Deep Sky Survey (UKIDSS). The UKIDSS project is defined in Lawrence et al (2007). UKIDSS uses the UKIRT Wide Field Camera (Casali et al, 2007). The photometric system is described in Hewett et al.~(2006), and the calibration is described in Hodgkin et al.~(2009). The pipeline processing and science archive are described in M. Irwin et al (in preparation) and Hambly et al (2008).  The Lockman Hole is contained within the Deep Extragalactic Survey and reaches a depth of $K$=21.  As it is located near the edge of the survey,  SpARCS1049+56 is only covered by the $K$-band.

\subsubsection{New CFHT  Imaging of the Lockman Field}

As part of SpARCS follow-up imaging,  35 square degrees overlapping the {\it Spitzer} Adaptation of the Red-sequence Cluster Survey Northern Fields were observed in the ugrz$^\prime$ bands using MegaCam on the 3.6 m CFHT, including the Lockman field (A. Tudorica et al. in preparation). Exposure times were 4201, 2291, 3007, and 6001 s for the $urgz^\prime$ filters respectively, and seeing averaged 0.8 arcsec. The data were reduced  using the ELIXIR and THELI pipelines. The absolute photometric calibration is based on the SDSS DR10, which was used to determine the field-by-field zeropoint values. The 5$\sigma$ limiting magnitudes are $u = 25.6 mag, g = 25.3 mag, r= 25.0 mag, z'=24.4 mag$.  For this work these data are used for display purposes only, in Figure 2. 

\subsection{New Imaging of SpARCS1049+56}

\subsubsection{JCMT SCUBA2 450/850$\mu$m }

We obtained 3 hours of integration in Band-1/2 weather  with the new SCUBA2 bolometer array on the JCMT.  SCUBA2 simultaneously observes at 450$\mu$m and 850$\mu$m with 8$\arcsec$ and 15$\arcsec$ FWHM beams respectively.  We employed the simple Daisy scan strategy which provides uniform sensitivity over the center $\sim$3$\arcmin$ area.  The maps were produced using the standard SCUBA2 data reduction pipeline (DRP) \citep{jenness11,geach13,chapin13} and beam-filtered to maximize sensitivity to point-sources.   In the central region of each map  (located at the cluster core) we reach an rms of $\sim$ 8 mJy beam$^{-1}$ at 450$\mu$m and  $\sim$ 1mJy beam$^{-1}$ at 850$\mu$m. 

\subsubsection{HST WFC3 Imaging}

We obtained infrared imaging of the cluster center with the F160W and F105W filters in Cycle 22 \footnote{Based on observations made with the NASA/ESA Hubble Space Telescope, obtained [from the Data Archive] at the Space Telescope Science Institute, which is operated by the Association of Universities for Research in Astronomy, Inc., under NASA contract NAS 5-26555. These observations are associated with program \# GO-13677 and \# GO-13747}.   These observations were obtained over nine separate visits, with total exposure times of 9237s and 8543s in F160W and F105W respectively.   

The data were processed using standard HST data reduction techniques, with some minor 
modifications. To combine and re-sample all images, we use the standard Space Telescope 
Science Institute (STSci) AstroDrizzle software from the DrizzlePac package 
\footnote{http://drizzlepac.stsci.edu/}, using the ``imedian'' combine type with $0\overset{''}{.}09$ pixels. The DrizzlePac software accounts for all distortion corrections, 
and also accounts for other instrument/detector specific calibrations, much of which is
contained in the image data quality array. Cosmic rays are flagged and 
corrected in the up-the-ramp sampling, and our 3-dither pattern allows unflagged 
hot pixels or uncorrected cosmic rays to be found and masked by AstroDrizzle.

To align the images, we detect objects in each image with SEP 
\footnote{https://github.com/kbarbary/sep}, and then apply standard point-set registration 
techniques to shift and rotate the images. We also use SEP to perform spatially 
variable background subtraction, masking the bright objects in the image to avoid 
bias in this background estimate.

\subsection{Spectroscopy}

\subsubsection{Keck-MOSFIRE Spectroscopy}

\textit{H}-band spectroscopy was obtained in 2013/14 using the near-infrared multi-object spectrograph MOSFIRE on the W.M.~Keck I Telescope \citep{mclean12}. The MOSFIRE \textit{H}-band filter is centered at $1.632 \mu m$, and has a spectral resolution $R=3660$. This allows the resolved observation of H$\alpha$ and the [N$_\mathrm{II}$] doublet, $\lambda\lambda6583,\, 6548$  at the redshift of SpARCS1049+56. Four separate \textit{H}-band masks were obtained with $\sim40$ slits per mask. During observations, each individual exposure was $120$s in length, with total integration times 45-60 minutes per mask. All observations were taken utilizing an ABBA dither pattern with a nod amplitude of $1\overset{''}{.}5$ and standard $0\overset{''}{.}7$ slit width.

We utilized the MOSFIRE data reduction pipeline (DRP v20140610) for the first step in data reduction. The DRP produces a two-dimensional spectrum and a corresponding sigma image from the raw flat and data frames for each object on the slit mask. All spectrum and sigma images are a sky-subtracted, wavelength calibrated, then registered to produce a stack of the A and B dither positions. All raw images were visually inspected for integrity before ingested into the DRP; any anomalous frames were thrown out. One step in the DRP involves interactively fitting the wavelength solution based on the sky lines during which we attempt to fit the wavelength solution to have a maximum residual difference of $0.2$ pixels for any one skyline. 

We then extracted  one-dimensional spectra from the two-dimensional spectra and sigma images following the extraction method of \citet{horne86}. Full details will be present in a future paper by DeGroot et al.~(2015, in preparation) but we give a brief outline here. Typical extraction routines simply sum the the sky-subtracted image data from the two-dimensional spectrum over a range that encloses the object's spectrum in the spatial direction. Our extraction applies a non-uniform pixel weighting, nominally gaussian in shape, to the extraction sum which increases the signal-to-noise of the resulting one-dimensional spectra. Redshifts are determined from developed line fitting routines that simultaneously fit H$\alpha$ and [N$_\mathrm{II}$] doublet emission lines as well as a linear continuum component via a Markov Chain Monte Carlo (MCMC) approach based on bayesian inference. Redshift errors are determined from the addition in quadrature of the random error from the MCMC fit and the systematic error from the interactive wavelength fitting step of the DRP.

\subsubsection{Spitzer IRS Spectroscopy} 

The position of the galaxy that we  identify  as the central BCG (Section 4.2) was serendipitously observed as part of an  earlier {\it Spitzer}-IRS survey of \citet{farrah08} (see Section 3.3 for further discussion of the identification of the BCG with this observation).  This program targeted infrared-bright systems (S$_{24{\mu}m} >$ 500mJy)  with infrared colors indicative of high-redshift galaxies.  The data acquisition, reduction and analysis  are described in Farrah et al., and we use these measurements in Section 3.3.  We further downloaded the spectrum from the Cornell Atlas of {\it Spitzer} IRS Spectra (CASSIS) which has developed its own analysis pipeline \citep{lebouteiller11}.

\section{Analysis}
\subsection{Discovery and Confirmation of the $\lowercase{z}=1.7$ SpARCS1049+56 cluster}

The northern/equatorial SpARCS SWIRE images cover 9.8, 4.5, 11.6 and 9.4 deg$^2$  (for a total of  $\sim$ 35 deg$^2$) over the ELAIS-N1, ELAIS-N2, Lockman and XMM-LSS  multi-wavelength survey fields respectively \citep{muzzin09}.  SpARCS1049+56 is located within the Lockman field (the largest single field) and is the second  richest cluster detection above $z=1.7$ within the 35 square degrees.   Unfortunately, it lies close to a chip gap on the $z'$ image and the outskirts of the northern half of the cluster lack coverage in this band. 

\begin{figure}
\includegraphics[scale=0.5]{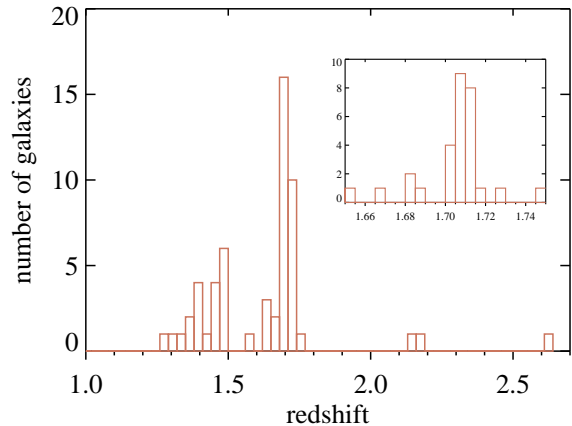}
\centering
\caption{The distribution of redshifts from the Keck-MOSFIRE spectroscopy. The inset shows a zoom in on the redshift    peak of $z=1.7089$.  A velocity cut of $\pm$1500 km/s of the mean cluster redshift results in 27 members within the 6$\times$3 arcmin field of view of MOSFIRE. \label{histogram}}
\end{figure}
\begin{figure*}
\centering

\hspace{-0.cm}\includegraphics[scale=0.8]{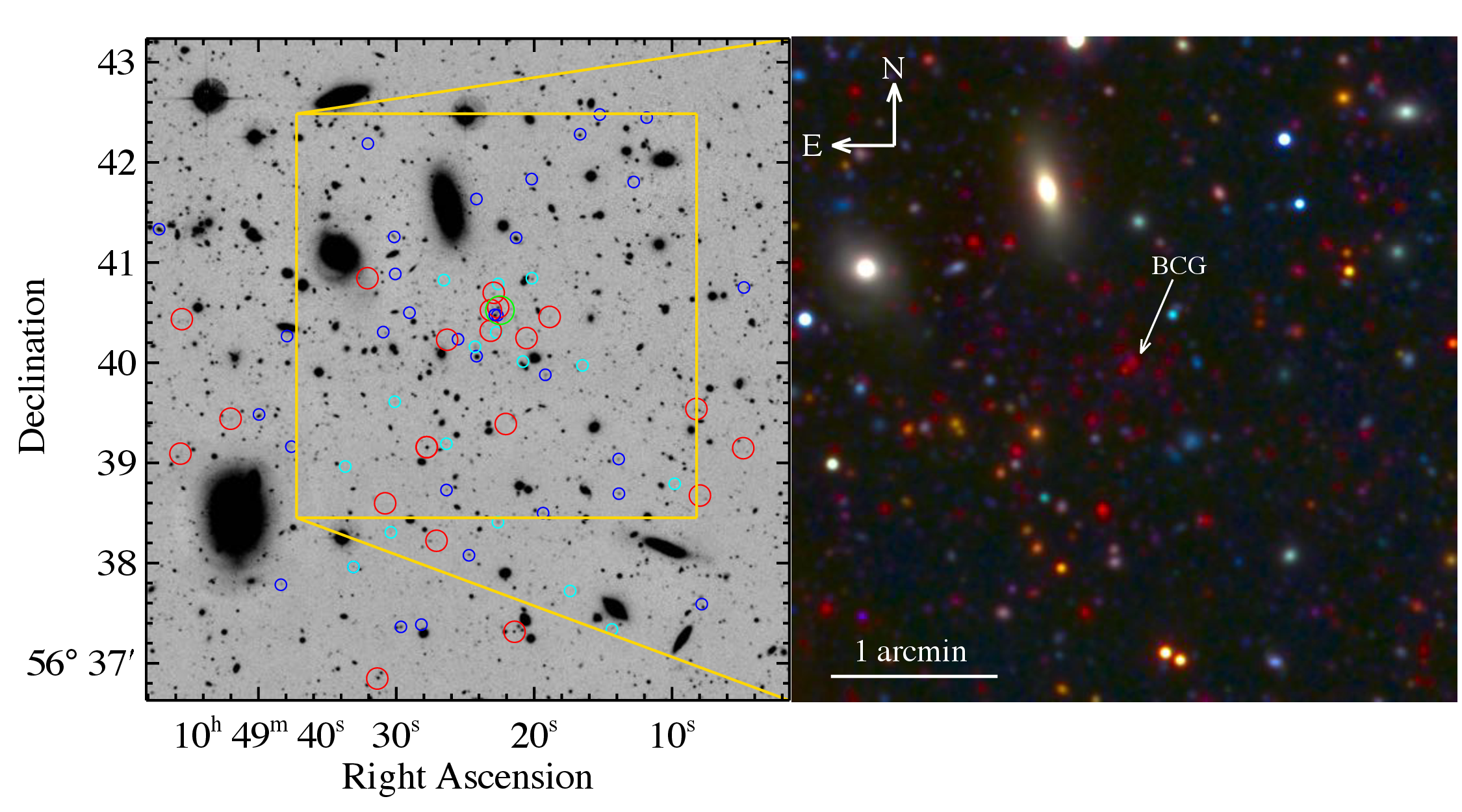}
\caption{{\bf Left:} The $K-$band image (greyscale) of the $\sim$6$\times$6 arcmin region around SpARCS1049+56.  Circles denote galaxies with spectroscopy from Keck-MOSFIRE. Large red circles indicate confirmed cluster members; smaller dark blue circles show foreground/background objects; and smaller cyan circles correspond to galaxies for which no redshift could be obtained.  This latter group contains foreground/background systems as well as cluster members with no strong emission lines.  The central green circle shows the location of the BCG, although the BCG itself is not easily visible in the this image (as it is very faint and red).  {\bf Right:}  The $ur3.6{\mu}m$ image of the cluster region with the location of the BCG again indicated.   \label{color}} 
\end{figure*}

The Keck-MOSFIRE spectroscopy  confirms the presence of a galaxy over-density at $z=1.7$.  An  average around the $z\sim$ 1.7 peak produces a cluster redshift of $z=1.7089$.   
 In Figure \ref{histogram} we show the distribution of 70 redshifts measured from four MOSFIRE masks of 146 independent  slits, over a field-of-view of $\sim$ 6$\times$3 arcmin, and in Figure \ref{color} we show the physical locations of the galaxies with spectra.  The distribution of confirmed cluster members  (see discussion below) is spatially extended over several Mpcs, but  this is, in part, a consequence of our observing constraints rather than a reflection of the true spatial distribution. 
The MOSFIRE design prohibits complete sampling of the core, due to a required minimum separation between slits,  and thus the confirmed members are spatially spread out (Figure \ref{color}). Moreover, redshifts were obtained through the detection of H$\alpha$ and this could impose a radial bias in  the redshift success, since completely passive galaxies (which may dominate within the core) will not show emission lines.   

We estimate the cluster richness from the 3.6$\mu$m (rest-frame $\sim$J) data to be $N_\mathrm{gal}=$30 $\pm$ 8. We use the following definition of    $N_\mathrm{gal}$:  the number of background-subtracted galaxies within a 500kpc  radius of the cluster center  brighter than $M^*$+1 based on a passive evolving luminosity function with $z_{formation}$ = 10 \citep{muzzin07}.  
Applying the  conversion of  \citet{andreon14b}  this richness yields a  mass  within 500 kpc of  (3.8$\pm$1.2)$\times$10$^{14}$$M_\odot$, where the uncertainty does not include redshift evolution in this relation. We note that in a richness-limited sample, such as this, there may be a bias toward systems with a high richness-to-mass relation. Confirmation of this mass will require a gravitational lensing measurement or detection of the intracluster medium (ICM).  Currently  there is no X-ray coverage of this object, and it is not expected (given its estimated mass) to appear as an SZ decrement in the Planck all-sky map \citep{ade14}.   

  We classify as cluster members  galaxies which lie within $\pm$1500 km/s of the mean cluster redshift.   This provides a sample of 27 cluster members within a cluster-centric radius of 1.8 Mpc.  A standard biweight-mean yields a velocity dispersion of 430$^{+80}_{-100}$ km/s.   We use  the model calibrated relation between cluster viral mass and velocity dispersion to convert this to a virial mass of 8$\pm2 \times$10$^{13}$ $M_\odot$.
However, \citet{saro13}  have found that calculating the velocity dispersion using at least the brightest 30 red sequence galaxies within the virial radius is required to obtain the virial mass of a cluster to within 40\%, whereas we have used 27 H$\alpha$ emitters over a wider area. This region of the spectrum is also plagued by sky lines and does not allow us to fully, or symmetrically sample the redshift distribution.   We therefore attach an uncertainty of at least 40\% to this mass, or 8$\pm$ 3$ \times$10$^{13}$ $M_\odot$. 

Although the mass is uncertain, we can add this system to the growing list of clusters and massive groups known around or above $z = 1.7$.   The most massive published clusters are the \citet{stanford12} system at $z=1.75$  (IDCSJ1426+3508) with seven spectroscopically confirmed members and  X-ray and SZ detections   that  indicate a mass estimate of $M_{200} \sim $5$\times$10$^{14}$ $M_\odot$ \citep{brodwin12}, and the \citet{tozzi15} cluster with $M_{500}=$3.2$^{+0.9}_{-0.6} \times$10$^{14}$$M_\odot$.    JKCS041 \citep{andreon09,andreon14a,newman14} lies at $z=$1.803, with a  mass constrained by optical richness and X-ray properties to be   1.6$\times$10$^{14} $$M_\odot$.   The \citet{zeimann12} system is at a slightly higher redshift of $z=1.89$, but with no mass estimate reported. Finally, CL1449+0586 \citep{gobat13} is  confirmed at $z=2.0$ with  26 spectroscopic members. Its X-ray flux indicates  a mass of  5$\times$10$^{13}$$M_\odot$. With a mass of 0.8-3.8 $\times$10$^{14}$ $M_\odot$ SpARCS1049 is similar to these objects.



 \subsection{The Galaxy Population and the  Brightest Cluster Galaxy}
\begin{figure}
\centering
\includegraphics[scale=0.5]{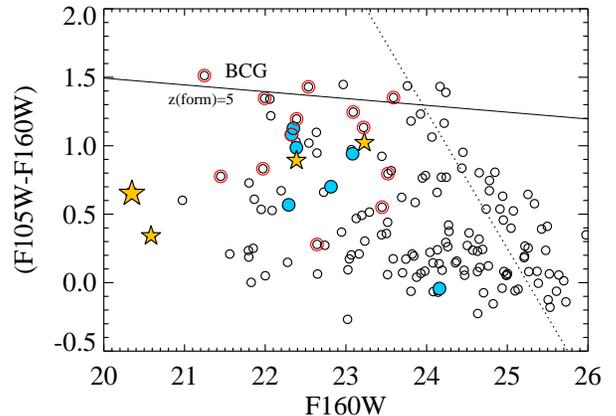}
\caption{ The HST NIR color-magnitude diagram, for all objects within  40$\arcsec$ of the cluster center.   For clarity we do not include photometric uncertainties, but show the $\sim$5$\sigma$  color-limit set by the F105W limit (diagonal dotted line); the same limit for F160W is $\sim$ 25.5 mag.   Small open circles correspond to all objects with photometric measurements in the inner 40$''$ radius of the HST image.  Solid orange stars  indicate confirmed non-cluster members (generally foreground), with the largest star corresponding to the bright blue line-of-sight companion to the BCG (J.S.~Huang, private communication).  Larger red circles  highlight  galaxies for which a  spectrum  was obtained, but no redshift was determined  because of a lack of spectral features. These are a combination of fore/background galaxies, and galaxies within the cluster with no strong emission lines (passive systems).  Finally, the large, solid blue circles indicate cluster members confirmed through emission-line redshifts.   We also include a model of the Red Sequence for a formation redshift of $z_\mathrm{form} = $ 5. The identified BCG (labeled) stands out as the brightest galaxy on the red-sequence.  \label{cmd}  }
\end{figure}

 Figure \ref{cmd} presents the color-magnitude diagram for SpARCS1049+56 using HST-WFC3 NIR filters which bracket the 4000A-break in the rest-frame.   We show all objects detected in both filters within the central 40$\arcsec$ ($\sim$ 320 kpc) radius of the HST image.   Orange stars show galaxies confirmed as foreground or background systems, with the largest star corresponding to a low-redshift  line-of-sight companion to the BCG (J.S.~Huang, private communication).  Solid blue circles  indicate cluster members confirmed through emission-line spectroscopy.  Large open red circles correspond to galaxies for which a spectrum was obtained, but no redshift was determined, due to a lack of features. These include fore/background systems and cluster members with no strong emission lines in the $H$-band, and emission-line cluster galaxies whose H$\alpha$ line is masked by a sky-line. 
 
An over-density of red-sequence galaxies is observed; this is expected since that is required for the initial cluster detection.  We show the location of the red-sequence (black line) predicted for a single stellar population (SSP) formed at $z_mathrm{form}=5$ \citep{bruzual03}.  It is interesting to note that all of the galaxies with colors consistent with the red sequence do not show evidence of strong line emission, implying the Red Sequence is already old and passive in this system. Nor are the identified cluster members very blue, rather they clump just below the red-sequence in the approximate region of the ``green-valley''.   This might be explained by the two primary biases of the MOSFIRE spectroscopic observations.  Targets were prioritized by $z^\prime  $  - 3.6$\mu$m color, to best select cluster members.  This biases the spectroscopy toward red galaxies.  However, redshifts required the detection of the H$\alpha$ emission line, thereby requiring star formation.  This combined selection could tilt the confirmed members  toward the reddest emission-line galaxies, or those in the green valley.   Still, the presence of these systems may indicate that a significant fraction of the galaxies within SpARCS1049 are transitioning from a star-forming episode to passive red-sequence galaxies. 

The BCG stands out clearly on the Red Sequence and is labeled in Figure \ref{cmd} with the next brightest system almost a full magnitude fainter.  H$\alpha$ was not observed for this galaxy, indicating it has a passively evolved stellar population (and see Section 3.3) with no measurable trace of star formation.  However, the IRAC colors indicate, through the presence of the stellar photospheric feature at rest-frame $1.6 \mu$m (the stellar bump), a photometric redshift of $z=1.7$.   

\subsection{Ultraluminous Infrared Emission Coincident with the Central Galaxy}

The center of the cluster is coincident with a previously known Ultraluminous Infrared Galaxy at $z=1.7$ in the Farrah et al.(2012) IRS  study of   extremely IR-bright high-redshift galaxies in SWIRE.   The sample was defined by  IRAC colors, to pre-select galaxies at $z\sim$ 1.7, and bright MIPS-24$\mu$m ($>500\mu$Jy) emission.  The target of the {\it Spitzer} spectroscopy was the BCG of the cluster and strong polycyclic aromatic hydrocarbon (PAH) features were detected, confirming a redshift of $z=1.7$.

Two other bright galaxies can be seen within the MIPS 24$\mu$m beam, to the SE.   Both of these galaxies have been spectroscopically confirmed to be foreground objects at $z=0.64$  (J.S.~Huang, private communication, the Multi-Mirror Telescope ) and $z=1.3$ (Keck-MOSFIRE),  for the blue and red galaxy respectively.  The detection of PAH lines at $z=1.7$ therefore removes these two galaxies as candidates for the MIPS emission.  Thus, the ground based-imaging leaves only the BCG as the candidate ID; although some diffuse emission is visible, no other source is seen within the 95\% confidence region. 

\begin{figure}
\includegraphics[scale=0.5]{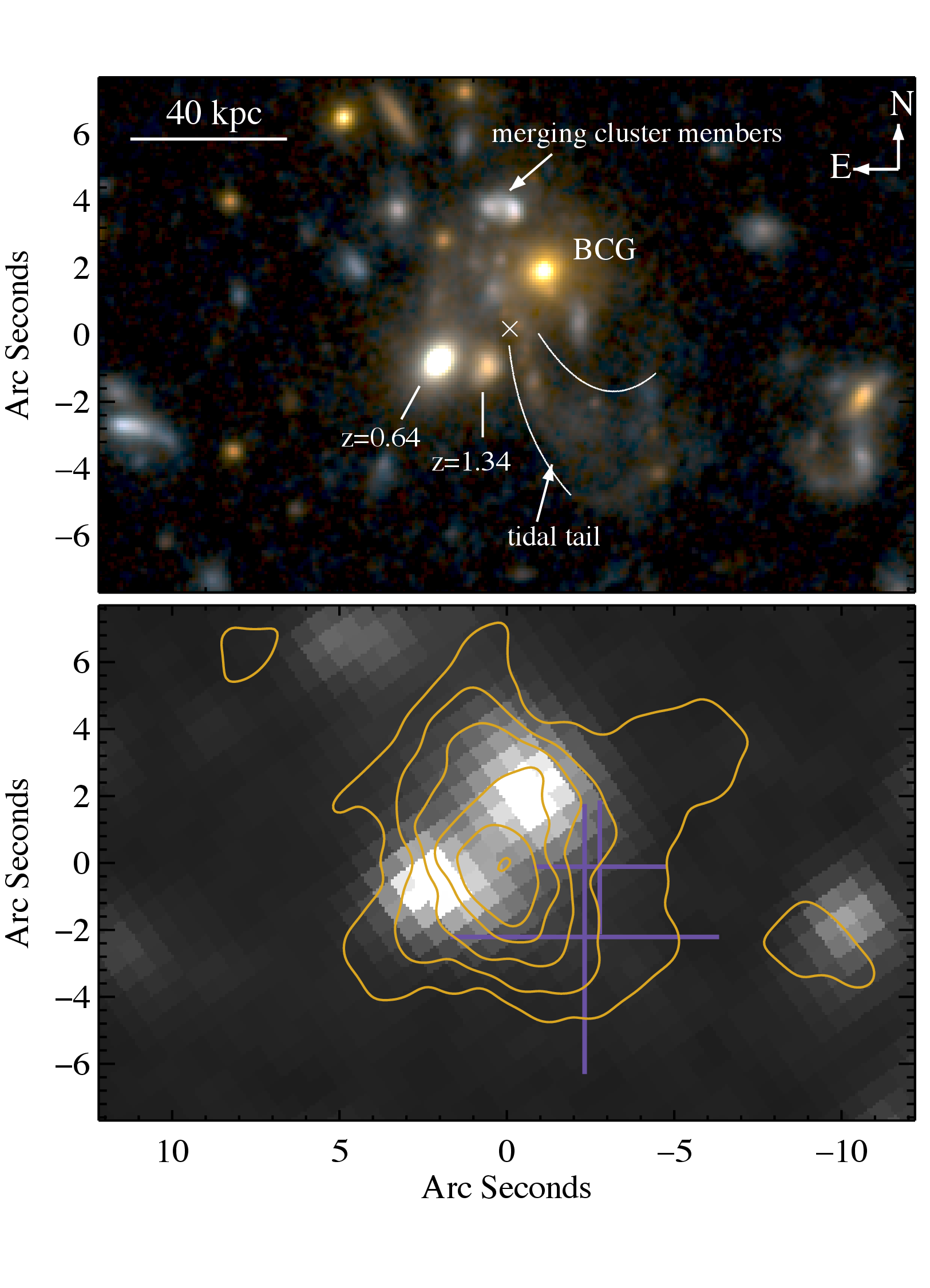}
\caption{{\bf Top:} the two-color (F160W, F105W) HST image of the central region of SpARCS1049, showing the complex morphology of the core.  Diffuse emission in a tidal-tail shaped structure is visible, and embedded along its length are multiple clumps. The cross marks the centroid of the 24$\mu$m emission.   {\bf Bottom:} The 3.6$\mu$m image of the same region (greyscale) with MIPS flux contours overlaid (orange) to show the location of the 24$\mu$m centroid and the mild extension of the source in the N/S direction.  The two purple crosses correspond to the 850$\mu$m (large) and 450$\mu$m (small) positions, where the size of the cross denotes 1/2 the FWHM of the beams.    \label{hst}}
\end{figure}

The deep HST  imaging reveals a more complicated situation, however. In Figure \ref{hst} we show the two-color composite image  of the central region using the F105W and F160W filters (rest-frame $\sim u'$ and $r'$, respectively).   Here  we see a long chain of small objects  
embedded within a diffuse filament or tidal-tail.     The MIPS centroid is located directly on top of one of these objects (in the approximate center of the chain) and thus it seems likely that the MIPS emission is coming not from the BCG itself, but from an object, or set of objects, within $\sim$16kpc of the BCG.  The details of this  structure are discussed  in the following section, but we note its morphology and the proximity to the BCG (possibly within its stellar envelope) suggest that the strong infrared emission is driven by an interaction with the central galaxy of the cluster.

\subsubsection{Constraining the Infrared Spectral Energy Distribution}

With such extensive multi-wavelength coverage of infrared-bright object, we are in a position to constrain its spectral energy distribution (SED) and derive simple parameters from it.  Flux measurements in the optical/NIR/MIR regimes are straightforward, however the large beam sizes of the {\it Herschel} instruments make confusion difficult to correct for in the crowded central regions of the cluster.  We therefore explain our method in detail here. 

 The SPIRE data  are publicly available and while there is emission coincident  with BCG in the HerMES 250$\mu$m catalog \citep{Wang13}, the increasing beam-size at 350 and 500$\mu$m results in a highly confused image of the cluster core.  However, utilizing positional priors from less-confused wavelengths can reduce the amount of source blending in far-infrared photometry.  In particular, \citet{Viero13} developed a novel algorithm called {\sc SIMSTACK} to stack SPIRE maps on prior positions from a user-defined catalog.  {\sc SIMSTACK} simultaneously stacks lists of potentially clustered populations, which effectively allows for a deconvolution of fluxes from multiple sources within each SPIRE beam.  We exploit the  \textit{Spitzer}-MIPS 24$\mu$m catalog from SWIRE for our positional priors.  

\begin{figure*}
 \includegraphics[scale=0.6]{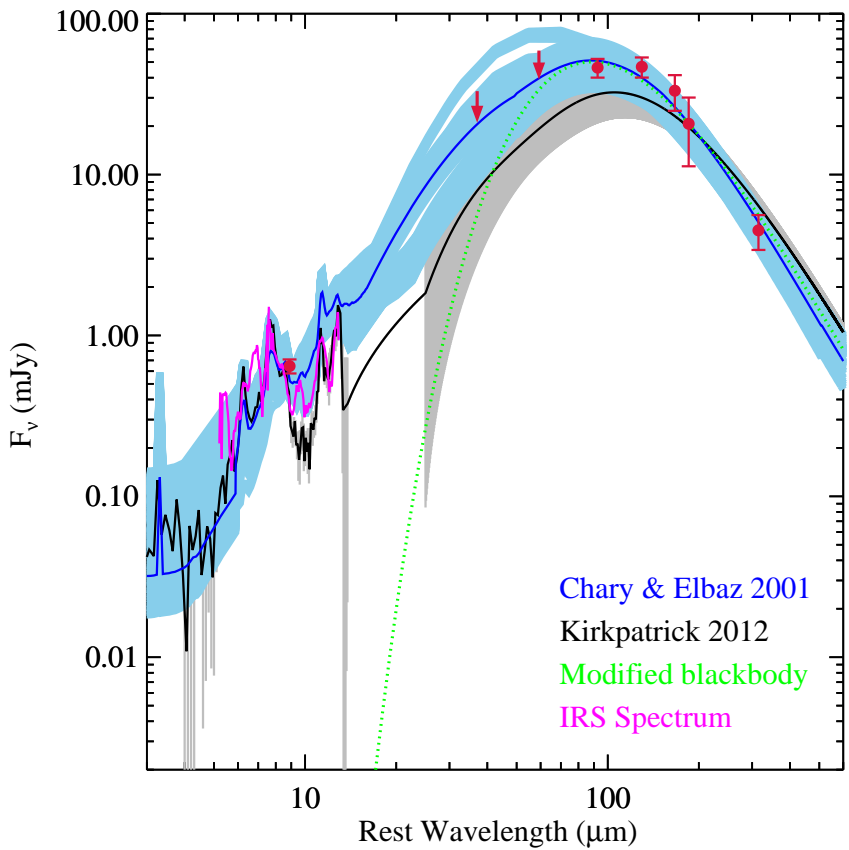}

\caption{The infrared spectral energy distribution from the combined MIPS/IRS/SPIRE/SCUBA2 flux measurements. The photometric points are shown in dark red  and the IRS spectrum in cyan.The red arrows show the 3$\sigma$ limits of the PACS measurements.  Overlaid (solid black line) is the  infrared luminous template SED for $z\sim$ 2 ULIRGs developed by \citet{kirkpatrick12}, with the scatter in this SED denoted by the grey shading. The blue curve corresponds to the best-fit SED from the \citep{charyelbaz01} template library, with the blue shading showing the  family of templates that produce a $<$3$\sigma$ fit.  
 Both SEDs are scaled to best-fit the photometric points only;  the fits do not include the IRS spectrum which is simply over-plotted as observed. The Chary \& Elbaz SED (without the optical points) yields the lowest $\chi^2$ and provides an infrared luminosity of $L_\mathrm{IR}$ = 6.6$\times$10$^{12}$ $L_\odot$.     The green dotted line shows  a modified black body ($\beta $= 1.7) fit to the SPIRE and SCUBA2 photometry which provides yet another estimate of the infrared luminosity, that is independent of the assumed SED.  The dust temperature is found to be T $= 44 \pm 5$ K and  $L_\mathrm{IR} \sim$ 4$\times$10$^{12}$ $L_\odot$, in  in good agreement with that of the full SED. The parameters of the template fits are listed in Table 1.  \label{sed} }
\end{figure*}

As we are only interested in the flux of the infrared source , we alter the code slightly to suit our needs.  We collect all 24$\mu$m sources within 90 arcsec of the BCG, which corresponds to $\sim$50, 30, and 15 independent beams at 250, 350, and 500$\mu$m, respectively.  We separate each source into individual lists, and simultaneously fit each list for the best value, yielding a flux of $46\pm6.2$, $47\pm6.7$, and $21\pm9.4$ for ascending SPIRE wavelengths, including the nominal confusion noise \citep{Nguyen10}.  We compare these fluxes to those derived from the HerMES XID catalog that uses the 250$\mu$m source positions (which have a FWHM beam $\sim3$ times larger than 24$\mu$m) as priors for 350 and 500$\mu$m.  While the individual SPIRE fluxes can differ by $2-3\sigma$ between the two methods, the resulting best fit SED (see Figure \ref{sed} and below), and subsequent infrared luminosity, are robust, differing only by 1.3\%. 

At 850$\mu$m we detect a point source at an offset of 4$\arcsec$ from the BCG location with a significance of 4.2$\sigma$.     A 450$\mu$m peak is detected at the same location at  significance of 4.0$\sigma$ (Figure 5).  Both of these offsets are within the expected positional error radius for SCUBA2, given the signal-to-noise of the detection, and these measurements are included in Figure \ref{sed}.  No useful constraints are obtained from the MIPS 70 and 170$\mu$m imaging of SWIRE.

In Figure \ref{sed} we show the complete SED for the far-infrared source using all available data from 24 to 850$\mu$m (Table 2), including the IRS spectrum in cyan.    The reduced IRS spectrum was obtained through the CASSIS.      We have fit  two sets of template SEDs to the data:  we use the the composite SED of \citet{kirkpatrick12} of star forming ULIRGs at $z\sim$ 2, and the library of templates from  \citet{charyelbaz01}.  We fit only to the photometry, for consistency,  but include a discussion of the information contained within the IRS spectrum below.

The Kirkpatrick SED is derived from infrared-luminous objects of similar luminosities to this source, at a similar redshift of $z\sim$ 2.  As there is only one spectrum, we fit for the amplitude and then integrate under the infrared portion (8-1000$\mu$m) of the SED to obtain the total infrared luminosity.  Uncertainties on the luminosity are produced by fitting the extremes of the uncertainties/scatter on the composite template. 

 The Chary \& Elbaz templates, on the other hand, are derived from local galaxy templates and thus may not be appropriate for galaxies at $z \sim$ 2.  Indeed, many studies have shown that using the CE01 SEDs to infer the L$_\mathrm{IR}$ from a single  24$\mu$m measurement over estimates the luminosity beyond $z = 1.5$ \citep{murphy09,nordon12,rodighiero10}.   As explored by \citep{elbaz11}, however, this discrepancy is reduced when one considers the full range of CE01 templates, rather than pre-selecting a single template based on the observed 24$\mu$m luminosity, and when fitting broader photometric coverage.   We therefore adopt the following fitting method for these templates.  We fit each available template (105 total variations) to the data \citep{pope08}. We calculate uncertainties on the infrared luminosities using a Bayesian approach.  
We calculate the normalized chi-square probability distribution as a function of Chary \& Elbaz  template and amplitude, ranging from 0-50 in steps of 0.05.  Each value in this grid corresponds to an infrared luminosity, which can then be weighted by the normalized chi-square probability.  The final uncertainty is given by the standard deviation of the weighted mean of the infrared luminosity: $\sigma_{\rm LIR} = \sqrt{\langle LIR^2 \rangle - \langle LIR \rangle ^2}$.


In Table \ref{sfrtable} we summarize the results of these fits.  Formally, the best representation of the data is the CE01 SED    This  fit produces $L_\mathrm{IR}$ = 6.2$\pm$0.9$\times$10$^{12}$ $L_\odot$, with a $\chi_\nu^2 = 0.4$.  We note the uncertainty on this value is small because only a small set of CE01 SEDs accurately represent the data. The Kirkpatrick model  yields $L_\mathrm{IR}$=4.1$\pm$ 1.5$\times$10$^{12}$ $L_\odot$, with  $\chi_\nu^2 = 4.0$. As can be seen in Figure \ref{sed}, the Kirkpatrick model over-predicts the amount of long-wavelength emission in the Rayleigh-Jeans tail.   However, the integrated luminosities of the two fits are consistent with each other at a level of $\sim1\sigma$.

The LIR emission can be converted to a SFR by  applying the standard \citet{kennicutt98} law, under the assumption that the dust is heated entirely by the stellar population with a Salpeter IMF, and these values are also included in Table \ref{sfrtable}.   The CE01 fit yields a SFR of 1070$\pm$160 $M_\odot$yr$^{-1}$, while the Kirkpatrick fit produces 690$\pm$260 $M_\odot$yr$^{-1}$.  These estimates are in agreement with the upper-limit on the SFR placed by \citet{farrah08} of  $< 1200$ $M_\odot$yr$^{-1}$ from the non-detection of the 6.2 $\mu$m line.  

 If an AGN is contributing to the IR luminosity this estimate must be reduced, however the presence of the PAH features in the spectrum, and the turn-over of the IRAC fluxes (implying a lack of very hot dust),  suggest that an AGN does not dominate the luminosity.
To investigate this we decompose the IRS spectrum into a star forming and AGN components following the method of \citet{pope08}.  We find an AGN contribution to the MIR of 50\%, corresponding to an AGN contribution to the FIR luminosity of $\sim$20\%.  We therefore apply this downward correction to the SFR.  This yields, for the best-fit SED, a SFR of 860$\pm$130 $M_\odot$yr$^{-1}$. 

\begin{deluxetable*}{lcccc}
\tablewidth{0pt}
\tablecaption{Luminosity and Star Formation Rate Estimates for the Infrared-Bright Central Galaxy System \label{sfrtable} }
\tablehead{
\colhead{Method/SED} & \colhead{$\chi_\nu^2$} & \colhead{$L_\mathrm{IR}$ }  & \colhead{SFR  } & \colhead{AGN removed SFR\tablenotemark{a}}  \\
\colhead{} & \colhead{} & \colhead{(10$^{12}$ $L_\odot$)} & \colhead{ ($M_\odot$ yr$^{-1})$} & \colhead{($M_\odot$ yr$^{-1})$}}
\startdata
Chary \& Elbaz   & 0.4 & 6.2 $\pm$ 0.9 & 1070 $\pm$ 160  &  856 $\pm$ 128 \\
Kirkpatrick  & 4.0 & 4.1 $\pm$ 1.5 &  \phantom{1}690 $\pm$ 290 &  552 $\pm$ 232 \\
\hline 

\\
6.2$\mu$m PAH\tablenotemark{b} & ... & ... & $<$ 1200 & ... \\

\enddata
\tablenotetext{a}{Assuming 20\% of the FIR luminosity is produced by the AGN}
\tablenotetext{b} {\citet{farrah08}}

\end{deluxetable*}



\vskip 0.5cm

\begin{deluxetable*}{cccccc}
\tablewidth{0pt}
\tablecaption{Infrared Photometry Measurements \label{sfrtable} }
\tablehead{
\colhead{24$\mu$m  (mJy)} & \colhead{250$\mu$m (mJy)}  & \colhead{350$\mu$m (mJy)}  & \colhead{450$\mu$m (mJy) } & \colhead{500$\mu$m (mJy)} &  \colhead{850$\mu$m} (mJy)  \\
}
\startdata
0.645 $\pm$ 0.065 & 46.2 $\pm$ 6.2 & 46.8 $\pm$ 6.7 & 33.2 $\pm$ 8.3 & 20.7  $\pm$ 9.4  & 4.5 $\pm$ 1.1 \\ 

\enddata
\end{deluxetable*}

\subsection{HST-revealed Beads on a String}

Prior to the HST NIR data, the BCG was the best candidate for the 24$\mu$m emission as it appeared to be the only viable option within the search radius of the source; however, the high resolution of the HST image has revealed a very complex structure within the core of the cluster, and has detected numerous additional objects within the search area. A color zoom-in of the central region is shown in Figure \ref{hst}. Here we see a chain of small clumps arranged as ``beads on a string'' and covering a linear area of roughly 66 kpc.   The clumps are embedded within diffuse emission which has the morphology of a single large tidal tail.      This tail appears to begin 1$''$ east (8 kpc) of the BCG, within its stellar envelope, and forms a distinct backwards J-like structure, ending to the south-west of the BCG.  

The morphology of this structure is reminiscent of interacting galaxies.  In particular, it is similar to the long single tidal arms of  the ``shrimp'' galaxies at $z\sim$ 1.4 classified by \citet{elmegreen07} in the GEMS and GOODS fields. These systems appear to be higher redshift analogs of merging systems such as M51 or the Antennae.  They consist of a single large tail, defined by diffuse emission and composed of several distinct and regularly spaced clumps.  The scale of the structure in SpARCS1049+56 is  similar to these, with a linear extent of 66 kpc, a width of $\sim$5  kpc and clump spacing ranging from 1 to 2 kpc. 
The mips 24$\mu$m emission peaks on one of the beads in the center of the tail and appears to be slightly extended along the direction of the tail (Figure \ref{hst})

\section{Discussion}

\subsection{ Dissipative Merger Induced Formation of the BCG}

The unique feature of this cluster is its extremely active and infrared-luminous central object. We conclude, based on the presence of PAH features and its NIR colors,  that the IR emission is dominated by star formation rather than an AGN.    The best SED fit yields L$_\mathrm{IR}$ = 6.2$\pm 0.9 \times$ 10$^{12}$ $L_\odot$ and a corresponding AGN-corrected SFR of 860$\pm$130 $M_\odot$yr$^{-1}$.  
In this respect the system is  similar to the Phoenix BCG \citep{mcdonald12}  within a $z = 0.6$ SPT cluster, but SpARCS1049+56 is seen at a significantly earlier cosmological epoch.   In both of these systems a substantial amount of gas is being delivered to the cluster center to fuel this intense starburst. In the case of the Phoenix the gas is deposited by a rapidly cooling intracluster medium (ICM) \citep{mcdonald13}.  X-ray imaging of the Phoenix  cluster indicates it has a cool core  and deep optical and UV imaging and spectroscopy  has revealed a network of large-scale gas filaments apparently streaming into the BCG \citep{mcdonald13,mcdonald14}.  

Although no X-ray information exists for SpARCS1049, the NIR (rest-frame UV and optical) data indicate a very different scenario.  
Rather than a spider-web arrangement of filaments, the HST imaging shows a single tidal tail, with  clumps along its length.  While this does not rule out the deposition of gas by a cooling flow,  the morphology is more consistent with  starbursts induced by gas-rich mergers.

Though the phenomenon of  mergers producing ``beads on a string'' tidal tails is well-established {\bf to our knowledge} \citep{elmegreen96} SpARCS1049 is only the second example of such an event in the center of a  galaxy cluster. 
Recently, a similar object was reported by \citet{tremblay14} in a lower-redshift ($z = $ 0.3) strong-lensing cluster, SDSSJ1531+3414.  This system is clearly associated with a major-merger 
consisting of the BCG and a massive companion. There are similarities between SpARCS1049 and SDSSJ1531+3414, namely the spacing of the beads ranges between 1 and 2.5 kpc and the linear extent is several tens of kpc long (66 kpc and 27 kpc respectively). However, there are some key differences as well. Firstly, a second massive galaxy, merging with the BCG is not  obvious in SpARCS1049.  The tidal tail appears to begin within the stellar envelope of the BCG with a smaller merging system that is spectroscopically confirmed to be within the cluster sitting at its tip.  The tail may be associated with this system which itself is interacting with the BCG, but does not have a comparable mass.    Secondly, although the rest-frame filters are necessarily different the beads/clumps of SpARCS1049 do not appear uniform in color, unlike those of SDSSJ1531+3414.  The clumps near the BCG-end of the tail are blue and similar in color to the merging pair at the tip, while the extended end of the tail has a color more consistent with the BCG itself.  Given that the MIPS emission is coincident with this second half of the tail, this may be due to extreme dust extinction in this region.  Finally, the scale of the star formation in SpARCS1049+56  is extremely different.  The total SFR of SDSSJ1531+3414 (determined from H$\alpha$) is 5 $M_\odot$yr${-1}$,  approximately two orders of magnitude smaller than we measure for SpARCS1049.

Assuming that the stellar mass produced in the burst will eventually be incorporated into the BCG we can assess its importance 
through a rough estimate of the specific SFR. For this we simply use the total SFR as seen in the IR and the existing approximate mass of the BCG.   To estimate the stellar mass we employ the color dependent rest-frame K mass-to-light ratio (M/L$_K$) calculated by  \citet{bell03b}, using the F160W-F105W color as an approximate rest-frame {\it u-g}.  This yields stellar mass  of 3$\pm0.4\times$10$^{11}$ M$_\odot$, where the error reflects a 10\% uncertainty in the photometry (corrected to a Salpeter IMF \citep{raue12}). Note that in the rest-frame K galaxies have a weak dependence of M/L on color:  an uncertainty in color of $\pm$0.5 would lead to an additional uncertainty of $\pm$0.5$\times$10$^{11}$ $M_\odot$. 
Thus, although it is undergoing a very rapid starburst of 860$M_\odot$ yr$^{-1}$ ,  the specific SFR of the  SpARCS1049+56 BCG is a  moderate $\sim$ 3-5 Gyr$^{-1}$, in line with mass-extrapolated main sequence field galaxies at $z\sim$ 2 \citep{daddi07}.  This could imply that  the mechanism of gas deposition is similar to that of the field at this redshift (major gas-rich mergers), or that galaxy self-regulation operates in the same way to limit the star formation efficiency, regardless of the fuelling process.   Nevertheless, the inferred SFR of  860$M_\odot$ yr$^{-1}$ would double the stellar mass of the galaxy  within roughly 400 million years therefore necessitating a rapid quenching mechanism.  This is  a similar timescale to that expected for a cooling flow of 100-500 Myr \citep{sarazin86}, with subsequent shut-down by an emerging AGN, but is also consistent with the gas depletion timescales for ULIRG mergers \citep{kennicutt98}.

Regardless of the central fueling mechanism, this  is an  optically rich cluster caught in the process of forming a large amount of stellar mass within the halo of its  BCG. SpARCS1049+56 therefore  offers a new opportunity to study the interplay between the global cluster physics, be it the cooling of the  gas halo or the interaction between infalling galaxies and the ICM,  and the evolution of the massive central galaxy  within the system at  a time when such processes may be more important to building clusters than today.  The rapid growth of BCGs through large-scale cooling of the ICM is a key ingredient at early times in SAMs \citep[e.g.][]{delucia07}, but  the supporting observational evidence is slim, largely because of the dearth of high-redshift ($z > 1.5$) over-densities available for study.  Evidence is mounting that growth of BCGS  through dry mergers is an important process \citep[e.g.][]{lidman12}, but these studies have been limited to $z\sim$ 1. It may be that SpARCS1049+56 is the first example of a new process, wet mergers of gas-rich galaxies, occurring  in the core of a $z\sim$1.7 cluster.



\section{Conclusions}

This paper presents the discovery and first analysis of a rich galaxy cluster at $z=1.708$  with a remarkable star-forming central galaxy.  While much additional work must be done to characterize the cluster itself  we can presently make the following conclusions.

\begin{enumerate}
\item{We have spectroscopically confirmed, through the presence of 27 member galaxies, an over-density of galaxies at $z=1.7089$.}
\item{The 3.6$\mu$m photometric richness is measured to be  $N_\mathrm{gal}$=30$\pm$8. Applying standard, low-redshift, conversions, the mass (within 500kpc) of the cluster is estimated to be $\sim$ 3.8$\pm$1.2$\times$10$^{14}$$M_\odot$.  The velocity dispersion, calculated from all 27 members over within 1.8 Mpc of the BCG, is 430$^{+80}_{-100}$ km/s, implying a lower mass of 8$\pm$3$\times$10$^{13}$ $M_\odot$.}

\item{The Brightest Cluster Galaxy is coincident with an Ultra-luminous Infrared Galaxy with an estimated luminosity of L$_\mathrm{IR}$ = 6.2$\pm$0.9$\times$10$^{12}$ L$_\odot$. The presence of PAH lines in its infrared spectrum and its rest-frame NIR colours indicate this energy is dominated by star formation, with an AGN corrected rate of of $\sim$ 860$\pm$130 $M_\odot$ yr$^{-1}$. This rate necessitates rapid gas deposition into the BCG system.  }

\item{Deep HST NIR imaging has revealed  complex morphology within the core.   A 66 kpc diffuse structure, reminiscent of a tidal tail, appears to emanate from the BCG. This structure contains several ($>$10) clumps of objects along its length, aligned as ``beads on a string''.  The MIPS emission is coincident with a set of red beads in the center of the tail.   } 
\item{The overall arrangement of the system suggests that the star formation is being driven by gas-rich galaxy-galaxy interaction, possibly involving the BCG. If so, this would be a new example of BCG mass growth through wet-mergers at high redshift.} 
\end{enumerate}

\acknowledgments
Some of the data presented herein were obtained at the W.M. Keck Observatory, which is operated as a scientific partnership among the California Institute of Technology, the University of California and the National Aeronautics and Space Administration. The Observatory was made possible by the generous financial support of the W.M. Keck Foundation. 
The authors wish to recognize and acknowledge the very significant cultural role and reverence that the summit of Mauna Kea has always had within the indigenous Hawaiian community.  We are most fortunate to have the opportunity to conduct observations from this mountain.  This material is based upon work in part supported by the U.S. Department of Energy, Office of Science, Office of High Energy Physics, under contract number No. AC02-05CH11231. This work is based in part on observations made with the Spitzer Space Telescope, which is operated by the Jet Propulsion Laboratory, California Institute of Technology under a contract with NASA. LD acknowledges support from European Research Council Advanced Grant: cosmicism. TMAW acknowledges the support of an NSERC Discovery Grant. Financial support for this work was provided by NASA through programs GO-13306, GO-13677, GO-13747, GO-13845 \& GO-14327 from the Space Telescope Science Institute, which is operated by AURA, Inc., under NASA contract NAS 5-26555.

 {\it Facilities:} \facility{HST (WFC3)}, \facility{Spitzer},  \facility{JCMT}, \facility{Keck:I}, \facility{Herschel}, \facility{CFHT}, \facility{UKIRT}

\end{document}